\documentclass{PoS}

\usepackage{graphicx} 
\usepackage{subfig}
\usepackage{amsmath,amssymb,mathtools}
\title{Quantized first-order phase transition and two sets of critical end point in droplet quark matter}

\ShortTitle{Quantized first-order phase transition and two sets of critical end point}

\author{\speaker{Kun Xu}\\
        Institute of High Energy Physics, Chinese Academy of Sciences,\\
        School of Nuclear Science and Technology, University of Chinese Academy of Sciences\\
        E-mail: xukun@ihep.ac.cn}

\author{Mei Huang\\
        School of Nuclear Science and Technology, University of Chinese Academy of Sciences\\
        E-mail: huangmei@ucas.ac.cn}

\abstract{The finite-size effect on the chiral phase transition is investigated in the Nambu--Jona-Lasinio model. To take into account finite-size effects, momentum integrals are replaced by momentum summations. The ground state of quark matter at finite size is favored when applying the periodic spatial boundary condition for quarks. The zero-momentum contribution is taken into account in the periodic boundary condition, and its contribution becomes important when the system size is comparable with the pion wavelength. When the zero-mode contribution becomes dominant, the conventional first-order chiral phase transition at high baryon chemical potential splits into two first-order phase transitions in small system of quark matter, and two sets of critical end point show up in the temperature and chemical potential plane.  }

\FullConference{Corfu Summer Institute 2018 "School and Workshops on Elementary Particle Physics and Gravity"\\
		(CORFU2018)\\
		31 August - 28 September, 2018\\
		Corfu, Greece}

\begin{document}

\section{Introduction}
The properties of strongly interacting matter under extreme conditions are controlled by Quantum Chromodynamics (QCD), and QCD phase transitions and phase structure have attracted much attention in the past several decades. It is widely believed that the chiral symmetry will be restored at high temperature and density. At small chemical potential and high temperature, it is a smooth crossover shown by lattice QCD calculation \cite{Fodor:2001au,Aoki:2006we,Bazavov:2011nk}, while it is expected a first-order phase transition would occur and there exists a QCD critical end point (CEP) at high chemical potential and low temperature \cite{Asakawa:1989bq,Klevansky:1992qe,Halasz:1998qr,Stephanov:2007fk,Lu:2015naa}. Exploring the QCD phase diagram and searching for the QCD CEP in the temperature and chemical potential $(T,\mu)$ plane are two of the most important goals of heavy-ion collision experimental studies, including beam energy scan at RHIC \cite{Adamczyk:2013dal,Aggarwal:2010wy,Luo:2015ewa,Luo:2017faz}, as well as for the future accelerator facilities at Facility for Antiproton and Ion Research (FAIR) and Nuclotron-based Ion Collider Facility (NICA).

Fluctuations of the conserved charges are sensitive to the first-order phase transition \cite{Stephanov:1998dy,Stephanov:1999zu,Ejiri:2005wq,Li:2018ygx,Asakawa:2009aj,Stephanov:2008qz}, and the kurtosis of the net proton number fluctuations $\kappa \sigma^2$ measured in the most central Au+Au collisions at first phase of beam energy scan program (BES-I) at RHIC  \cite{Adamczyk:2013dal,Aggarwal:2010wy,Luo:2015ewa,Luo:2017faz} show a non-monotonic energy dependent behavior. This non-monotonic behavior may serve as a strong indication of the existence of CEP \cite{Li:2018ygx}. There are many effects that might be important to interpret  the experimental data, for example, centrality bin width correction\cite{Luo:2011ts,Luo:2013bmi}, efficiency correction for the cumulants and finite size effect. In a real QGP phase created in heavy ion collision, the system has a finite size \cite{PHENIX:2018lia}.

There are many works that have discussed the effects of finite size \cite{Kiriyama:2006uh,Braun:2011iz,Pan:2016ecs,Balian:1970fw,Berger:1986ps,Madsen:1994vp,Palhares:2009tf,Almasi:2016zqf,Braun:2005fj,He:1996fc,Kogut:1999um}, and different strategies have been developed to incorporate the finite size effects: using the multiple reflection expansion(MRE) \cite{Kiriyama:2006uh,Balian:1970fw,Berger:1986ps,Madsen:1994vp}, replacing momentum integrals  by momentum summations \cite{Braun:2011iz,Palhares:2009tf,Almasi:2016zqf,Braun:2005fj}, or giving a lower momentum cutoff in momentum integrals \cite{Pan:2016ecs}.
Like in lattice QCD doing numerical simulations on finite and discrete Euclidean space-time, the general method in finite size system is to replace the momentum integral with momentum summation. The most natural choice of spatial boundary condition for bosons is periodic, i.e. the momentum is summed from the exact zero-momentum mode. However, it has been always ambiguous for choosing the spatial boundary conditions for fermions or quarks. In literatures, both anti-periodic (AP) and periodic (P) spatial boundary conditions have been applied and neither has been strictly excluded.

The typical length in QCD systems is the pion wave-length $\lambda_{\pi}=1/m_{\pi}\sim 1.41 {\rm fm}$, when the system size $L$ is comparable with the pion wave-length $L\sim \lambda_{\pi}$, the size effect on the system will become significant, thus the boundary condition becomes important. Applying the anti-periodic and periodic spatial boundary conditions to quarks induces opposite results on vacuum properties:  the anti-periodic spatial boundary condition for quarks induces that the chiral symmetry restores in the small system, while the periodic spatial boundary condition induces the enhancement or catalysis of the chiral symmetry breaking in the vacuum. In most cases, the anti-periodic spatial boundary condition has been applied for quarks to keep the permutation symmetry between the time and space directions \cite{Almasi:2016zqf,Klein:2017shl}. Another reason of applying the anti-periodic boundary condition to quarks is to get consistent results of volume dependent pion mass from chiral perturbation theory (ChPT) \cite{Luscher:1985dn}, where the pion mass increases with the decrease of the system size.

This talk is based on our work in Ref. \cite{Xu:2019gia}. We will firstly compare the thermodynamical potential of the small system by applying the periodic and anti-periodic spatial boundary conditions for quarks in the framework of Nambu-Jona-Lasinio (NJL) model, and find that the ground state of the small system is favored by applying the periodic boundary condition where the zero-momentum mode contribution is taken into account. In Sec.III, we will show the results of catalysis of chiral symmetry breaking and the constant mass of pseudo NG pions in small systems by applying the periodic boundary condition. Then we will show the interesting result of quantized first-order phase transition and two sets of CEP observed in cold droplet quark matter in Sec.IV. At last we give summary and discussion.


\section{The ground state of droplet quark matter}

In this talk, we focus on discussing the boundary condition of quarks in QCD system, therefore we can neglect the finite size effect on gluon dynamics.
We take the simplest four-fermion interacting 2-flavor Nambu--Jona-Lasinio (NJL) model with only scalar interaction, and its Lagrangian density is given by \cite{Klevansky:1992qe}:
\begin{equation}
\mathcal{L}=\bar{\psi}(i\gamma^\mu \partial_\mu -m_0)\psi+G[(\bar{\psi}\psi)^2+(\bar{\psi}\gamma^5 \vec{\tau}\psi)^2].
\end{equation}
Where $\psi=(u,d)^T$ is the quark doublet with two light quark flavors, and the current mass is assumed to be equal $m_u=m_d=m_0$, $\vec{\tau}=(\tau^1,\tau^2,\tau^3)$ is the isospin Pauli matrix and $G$ is the coupling constant in the scalar channel. Introducing the auxiliary scalar and pseudo-scalar fields $\sigma$ and $\vec{\pi}$ and their condensations are defined as:
\begin{equation}
	\sigma=-2G \langle \bar{\psi}\psi \rangle,\quad \vec{\pi}=-2G \langle \bar{\psi}\gamma^5 \vec{\tau}\psi \rangle.
\end{equation}
Considering $\vec{\pi}=0$ in the vacuum and taking mean-field approximation, the thermodynamical potential of the NJL model takes the following form:
\begin{equation}
\Omega=\frac{(M-m_{0})^{2}}{4 G}-2 N_{c} N_{f} \int\frac{d^3 p}{(2\pi)^3}\left\lbrace E+T\ln(1+e^{-\frac{E+\mu}{T}})+T\ln(1+e^{-\frac{E-\mu}{T}})\right\rbrace ,
\end{equation}
where $N_c$ and $N_f$ are the number of colors and flavors, and $T,\mu$ the temperature and quark chemical potential, respectively. The quark quasiparticle energies $E$ and constituent quark masses $M$ are given by:
\begin{equation}
E=\sqrt{p^{2}+M^{2}},\quad  M=m_{0}+\sigma .
\end{equation}
The NJL model is a non-renormalized model, thus a regularization scheme is needed to avoid infinity. To obtain elegant result,  here we take the Pauli-Villars regularization scheme, then the effective potential has the form of:
\begin{equation}
\Omega=\frac{(M-m_{0})^{2}}{4 G}-2 N_{c} N_{f} \int^{\infty}_{-\infty}\frac{d^3 p}{(2\pi)^3} \left\lbrace   \sum_{j=0}^{3} c_j \sqrt{E^2 + j\Lambda^2}+T\ln(1+e^{-\frac{E+\mu_B}{T}})+T\ln(1+e^{-\frac{E-\mu_B}{T}}) \right\rbrace   ,
\end{equation}
where
\begin{equation}
c_0=1,\quad c_1=-3, \quad c_2=3, \quad c_3=-1
\end{equation}
 are fixed in the regularization framework. Other model parameters, e.g. $G$ and $\Lambda$ are fixed by the pion decay constant $f_{\pi} =93\text{MeV}$ and quark constituent mass $M=330\text{MeV}$ in the vacuum, and we fix $m_0=5.5\text{MeV}$, $N_c = 3$, $N_f = 2$. In order to find the minimum of potential $\Omega$, we need to solve the following gap equation:
\begin{equation}
\frac{\partial \Omega}{\partial \sigma}=0.
\label{gapequ}
\end{equation}

We now put quark matter in a cubic box with finite length $L$, and we replace the momentum integral with the summation of the discrete momentum:
\begin{equation}
\int \frac{d^3 p}{(2 \pi)^3}\rightarrow\frac{1}{V}\sum_{p},
\end{equation}
with $V=L^3$ the volume of the system. The effective potential of the quark matter in finite size now takes the form of:\\
\begin{equation}
\Omega=\frac{(M-m_{0})^{2}}{4 G}-\frac{2 N_{c} N_{f}}{V}\sum_{\vec{p}}\left\lbrace \sum_{j=0}^{3} c_j \sqrt{E^2 + j\Lambda^2}+T\ln(1+e^{-\frac{E+\mu}{T}})+T\ln(1+e^{-\frac{E-\mu}{T}})\right\rbrace .
\end{equation}

For fermions, in the time direction, only anti-periodic boundary condition is allowed, however, in the spatial direction, there is no strict rule to rule out either the periodic boundary condition
\begin{equation}
	\vec{p}^{2}=(\frac{2\pi}{L})^{2}\sum_{i=x,y,z}n_{i}^{2},
\end{equation}
or the anti-periodic boundary condition
\begin{equation}
	\vec{p}^{2}=(\frac{2\pi}{L})^{2}\sum_{i=x,y,z}(n_{i}+\frac{1}{2})^{2},
\end{equation}
with $n_i=0,1,2,...$ non-negative integers in both equations. Therefore both spatial boundary conditions have been applied for quarks for several decades in literatures \cite{Kiriyama:2006uh,Braun:2011iz,Pan:2016ecs,Balian:1970fw,Berger:1986ps,Madsen:1994vp,Palhares:2009tf,Almasi:2016zqf,Braun:2005fj,He:1996fc,Kogut:1999um}. The most important difference between these two boundary conditions is that whether zero-momentum mode contribution is taken into account: the zero-mode contribution is included in the periodic spatial boundary condition while it is not included in the anti-periodic boundary condition.

\begin{figure}
	\centering
\includegraphics[width=0.45\textwidth]{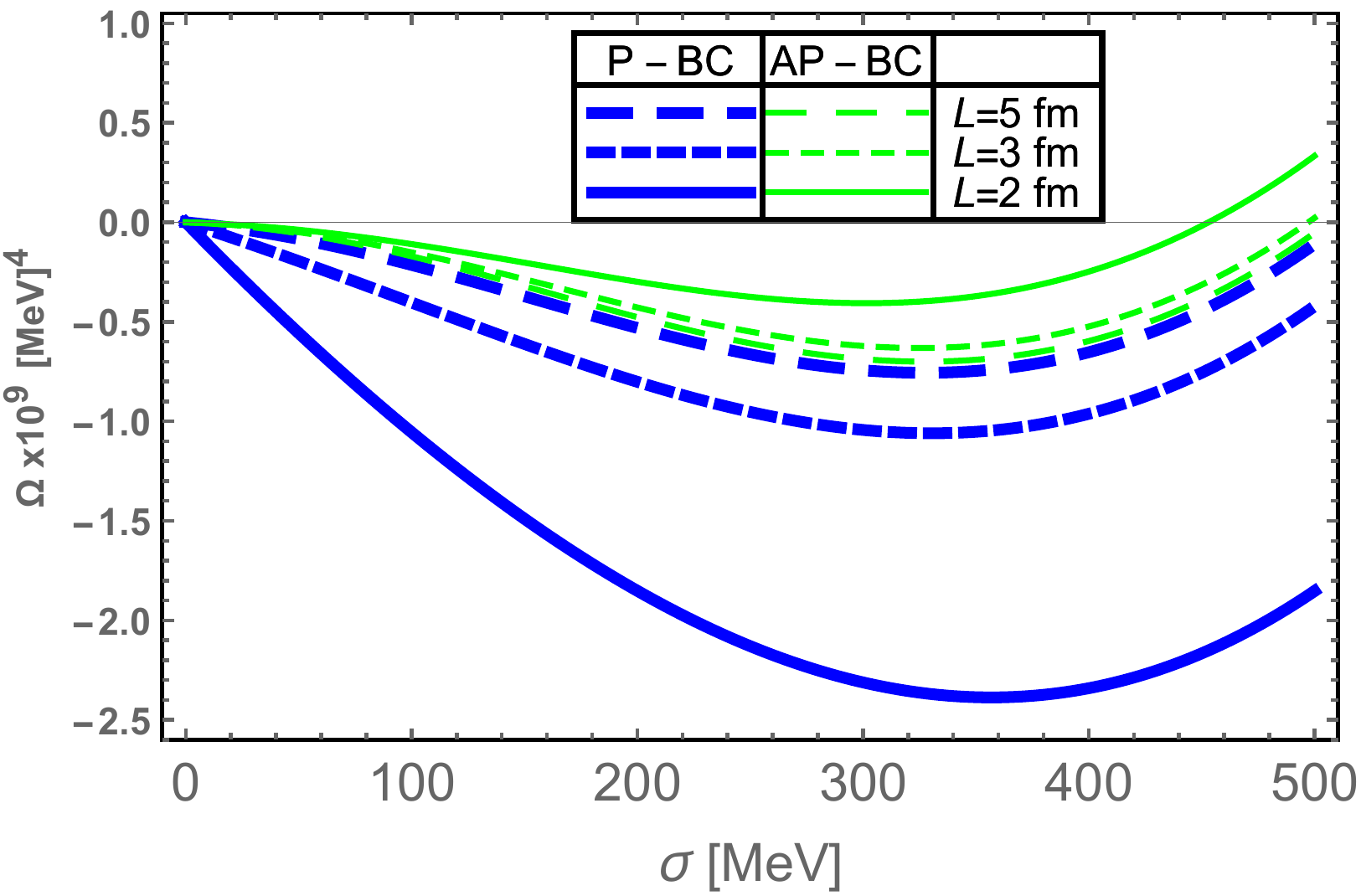}
\caption{The effective potential of the small system as a function of the chiral condensate $\sigma$ for three different sizes $L=5,3,2 ~ fm$ at $T=0,\mu=0$ by applying the periodic boundary condition (P-BC) (in blue lines) and the anti-periodic boundary condition (AP-BC) (in green lines), respectively.}
\label{fig:omega}
\end{figure}

In order to determine the ground state of the system, we compare the thermodynamical potential of quark matter in finite size in Fig.\ref{fig:omega} by applying the periodic and antiperiodic boundary conditions, respectively. At zero temperature and chemical potential $T=0,\mu=0$, it is observed that when applying the periodic spatial boundary condition for quarks, the effective potential as a function of the chiral condensate becomes lower with the decreasing of the size, while when applying the anti-periodic spatial boundary condition for quarks, the effective potential becomes higher when system size becomes smaller. Therefore, at fixed size, the thermodynamical potential by applying the periodic boundary condition is much lower than that by applying the anti-periodic boundary condition. This indicates that the finite size system prefers the periodic spatial boundary condition for quarks, in which the zero-momentum mode is taken into account.

\section{The catalysis of chiral symmetry breaking and the pseudo NG pions}

\begin{figure}
	\centering
\includegraphics[width=0.45\textwidth]{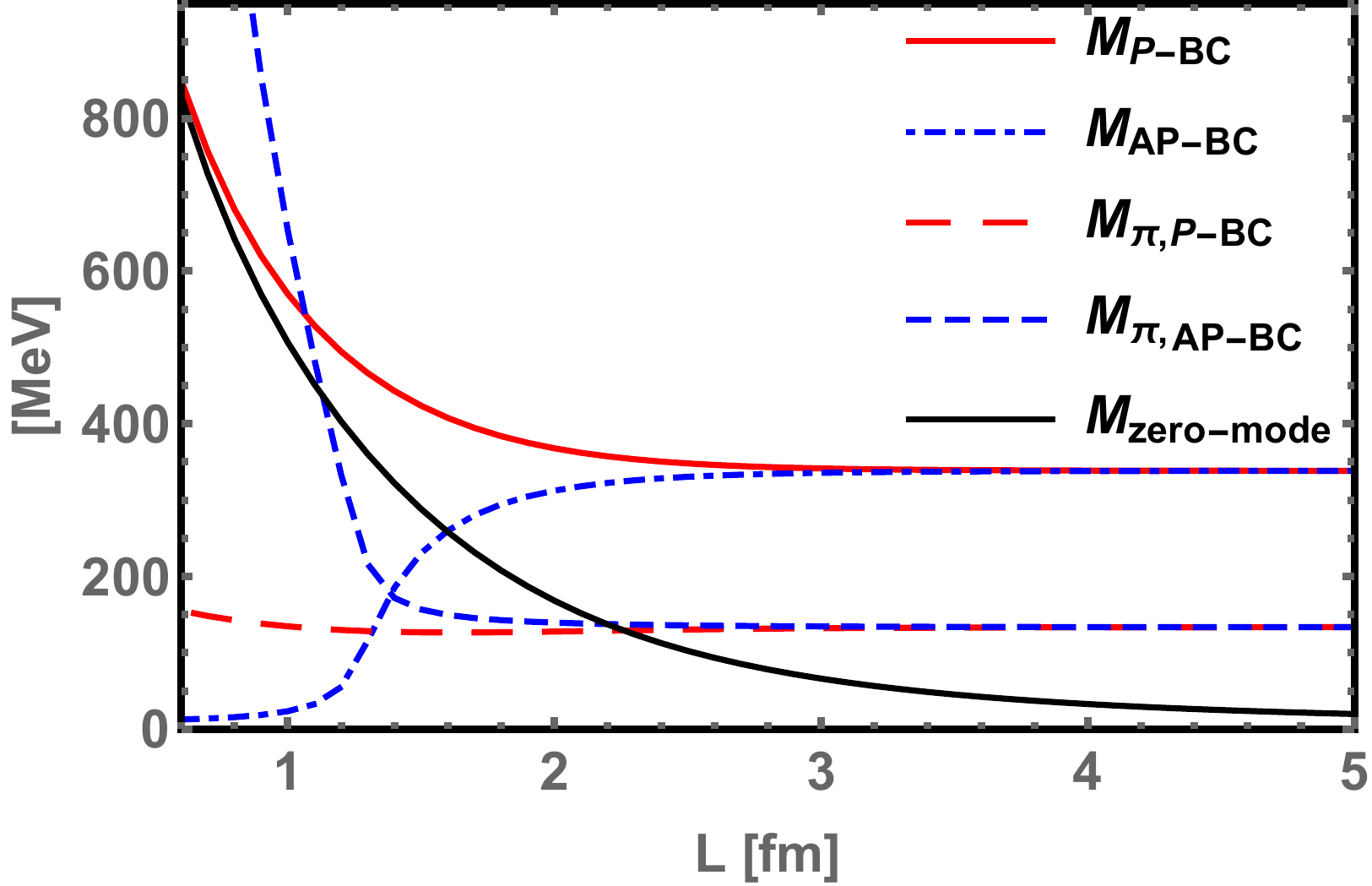}
\caption{The constituent quark mass $M$ and pion mass $M_{\pi}$ as a function of the system size $L$ at $T=0,\mu=0$ by applying the P-BC and AP-BC for quarks, respectively. $M_{\text{zero-mode}}$ is the constituent quark mass obtained with only zero-mode contribution. }
\label{fig:M-P-AP-PV}
\end{figure}

From the thermodynamical potential in Fig. \ref{fig:omega}, we can read the chiral condensation in the vacuum, which can also be solved from the gap equation Eq.(\ref{gapequ}), and the corresponding constituent quark mass is $M=m_0+\sigma$. In the NJL model, mesons are $\bar{q}q$ bound state or resonance, and can be obtained from the quark-antiquark scattering amplitude. The mesons are constructed by summing up infinite quark-loop chains in the random phase approximation(RPA) to the leading order of $1/N_c$ expansion \cite{Klevansky:1992qe}. The one-loop polarization function of pions takes the form of
 \begin{equation}
 \Pi_{\pi}(q)=-i\int\frac{d^4k}{(2\pi)^4}Tr[i\gamma_5\vec{\tau}S(k)i\gamma_5\vec{\tau}S(p)],
 \end{equation}
 with $k=p+q$. The pion mass is determined by the gap equation
 \begin{equation}
 \label{Gappi}
 1-2G\Pi_{\pi}(q^2=M_{\pi}^2)=0.
 \end{equation}

In Fig. \ref{fig:M-P-AP-PV} we show the constituent quark mass $M$ and pion mass $M_{\pi}$ at $T=0,\mu=0$ by applying the P-BC and AP-BC for quarks, respectively. It can be seen that when applying the periodic boundary condition for quarks, with the decreasing of the system size, the chiral condensate enhances, especially when $L<2 fm$, the chiral condensate enhances dramatically in small system. It is noticed that in the chiral symmetry breaking vacuum, pion mass is a constant and pions keep as the pseudo Nambu-Goldstone (NG) bosons. This is the familiar phenomenon of catalysis of chiral symmetry breaking, which has also been observed in quark matter under strong magnetic fields \cite{Gusynin:1994re,Miransky:2015ava}, where only neutral pion keeps as pseudo NG boson \cite{Liu:2018zag}.
We can understand the similarity between the small system and system under strong magnetic field. Remembering that the magnetic length $l$ for particle carrying charge $q$ is proportional to the inverse of the square root of magnetic field, i.e. $ l\sim \frac{1}{\sqrt{|q| B}}$ \cite{Tong:2016kpv}, in some sense putting charged particles under strong magnetic field is similar to put these particles in an elongated cylinder with small radius $l$. On the other hand, if the anti-periodic boundary condition is applied for quarks, it is found that the quark mass decreases and the chiral symmetry becomes restored in small system, and pion mass become much heavier in the vacuum.

As we mentioned earlier that the only difference between the periodic and anti-periodic boundary conditions is whether to take into account the zero-momentum mode contribution. For finite size system, the momentum becomes discrete, and the gap between the zero-mode and the first-mode is $\frac{2\pi}{L}$. When the system size $L<2 fm$, the zero-momentum mode contribution dominates, which can be read from the constituent quark mass $M_{\text{zero-mode}}$ only considering the zero-mode contribution as shown in Fig. \ref{fig:M-P-AP-PV}.

\section{Quantized first-order chiral phase transition}

As we have shown above that the ground state of the small system is favored when the periodic spatial boundary condition is applied to quarks, in the following, we will only choose the periodic boundary condition for quarks and investigate how the finite size affects the chiral phase transition.

\begin{figure}
	\centering
	\subfloat[$L=10fm$]{\includegraphics[width=200pt]{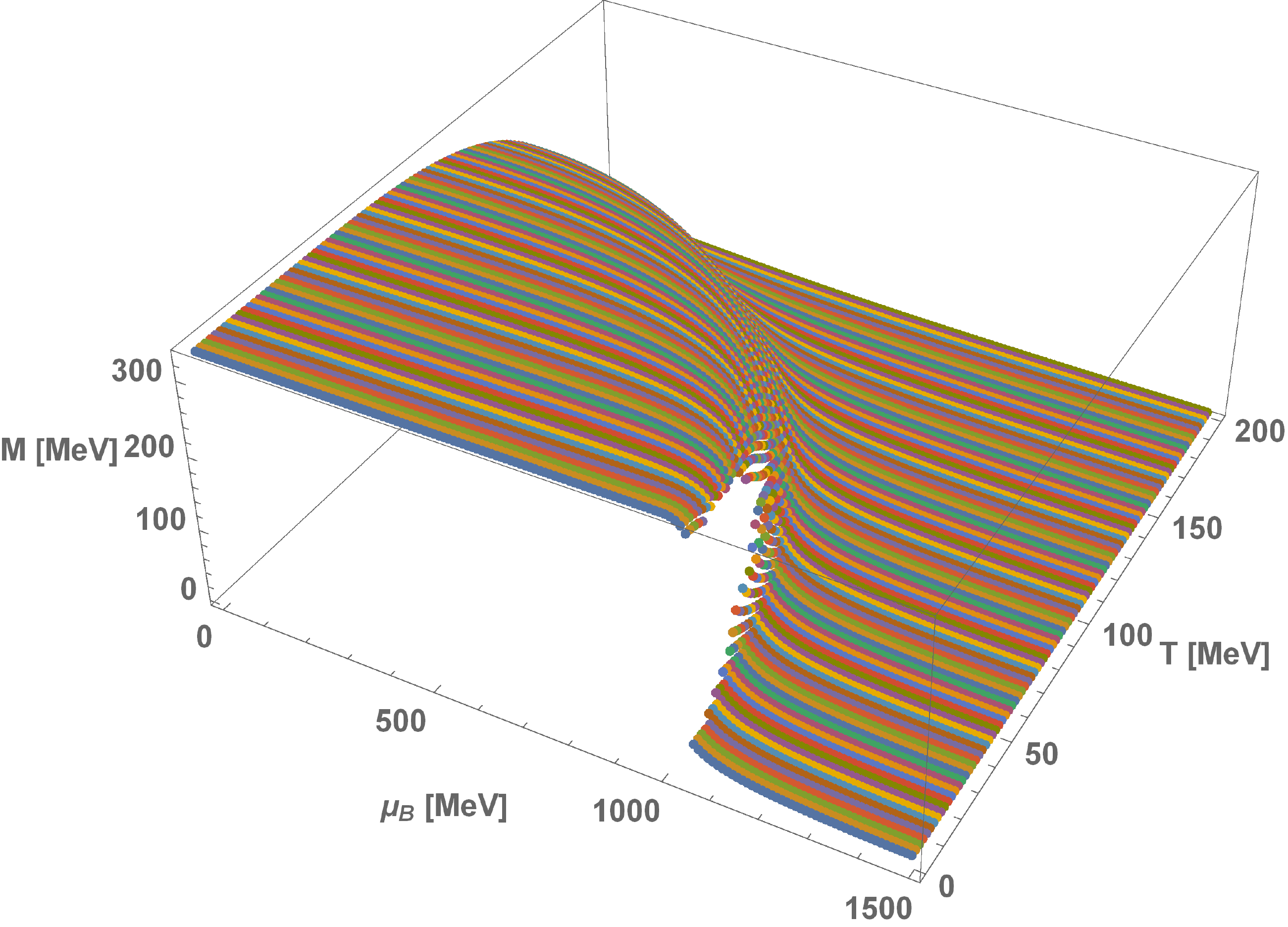}\label{subfig:l10}\hspace{20pt}}
	\subfloat[$L=5fm$]{\includegraphics[width=200pt]{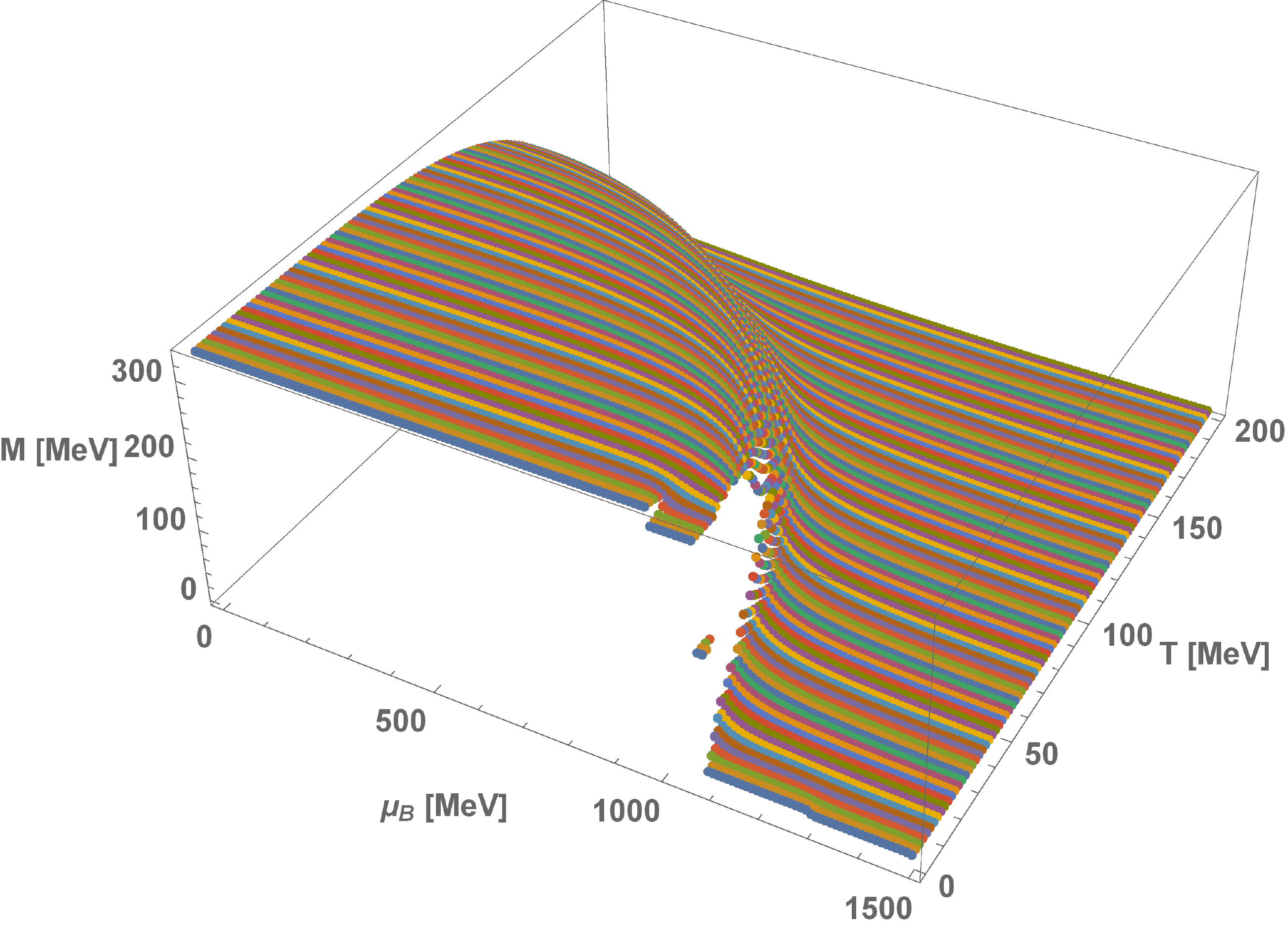}\label{subfig:l5}\hspace{20pt}}\\
	\subfloat[$L=3fm$]{\includegraphics[width=200pt]{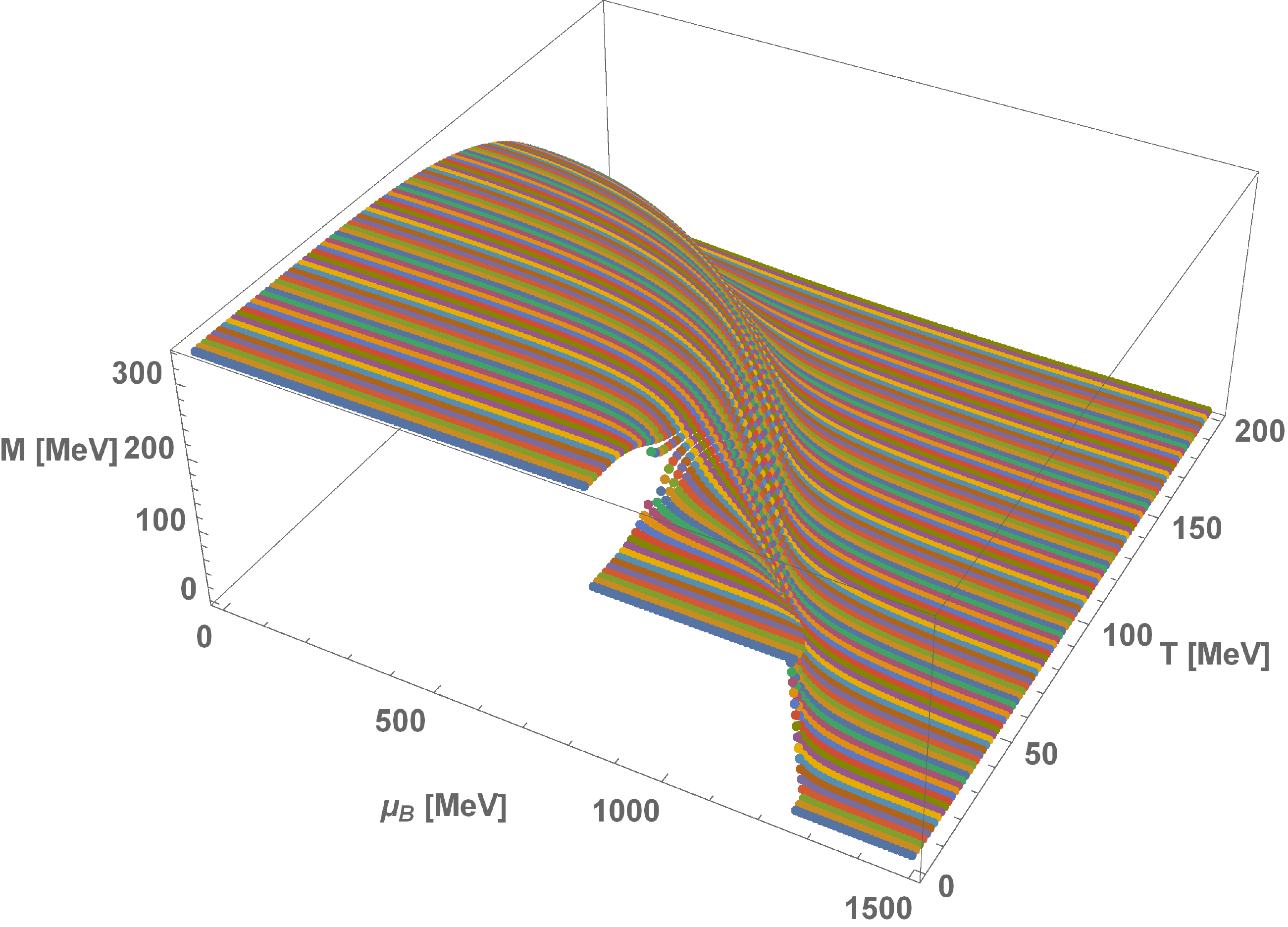}\label{subfig:l3}\hspace{20pt}}
	\subfloat[$L=2fm$]{\includegraphics[width=200pt]{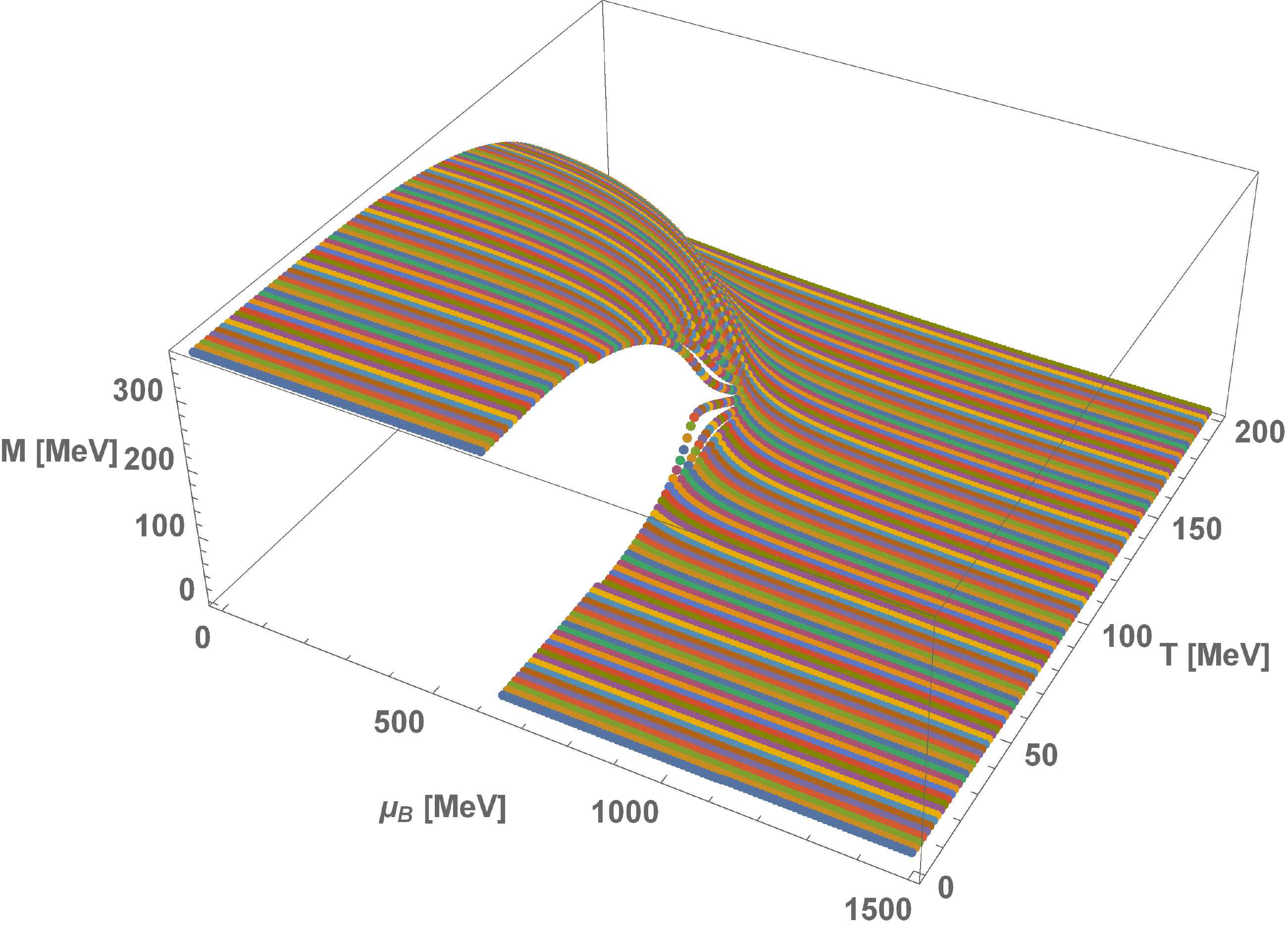}\label{subfig:l2}\hspace{20pt}}
	\caption{The constituent quark mass $M$ as functions of the temperature $T$ and baryon chemical potential $\mu_B$ for different sizes $L=10,5,3,2 fm$.}
\label{fig:quarkmass-mu-T-L}
\end{figure}

By solving the gap equation Eq.(\ref{gapequ}), the constituent quark mass $M$ as functions of the temperature and baryon chemical potential $\mu_B=3 \mu$ for different sizes $L=10,5,3,2 fm$ can be obtained and the results are shown in Fig.\ref{fig:quarkmass-mu-T-L}.

For the size $L=10 fm$, it is seen from Fig.\ref{subfig:l10} that the chiral phase transition in the $(T,\mu_B)$ plane looks almost the same as that in the case of $L=\infty$, and the phase transition is a smooth cross-over at high temperature,  and a first-order at high baryon chemical potential. However, when the size decreases, e.g. at $L=5,3 fm$, some structures show up in the low temperature and high baryon chemical potential region as shown in Fig.\ref{subfig:l5} and Fig.\ref{subfig:l3}, it is observed that there are two jumps and the first-order phase transition now splits into two first-order phase transition. For convenience, we mark the 1st jump at smaller chemical poential as "PT1" and the 2nd jump at larger chemical potential as "PT2".  When the size decreases, the second jump PT2 shrinks while the magnitude of the first jump PT1 gets larger. At last, PT2 vanishes and only PT1 survives in small size as shown in Fig.\ref{subfig:l2}.

\begin{figure}
	\centering
	\includegraphics[width=280pt]{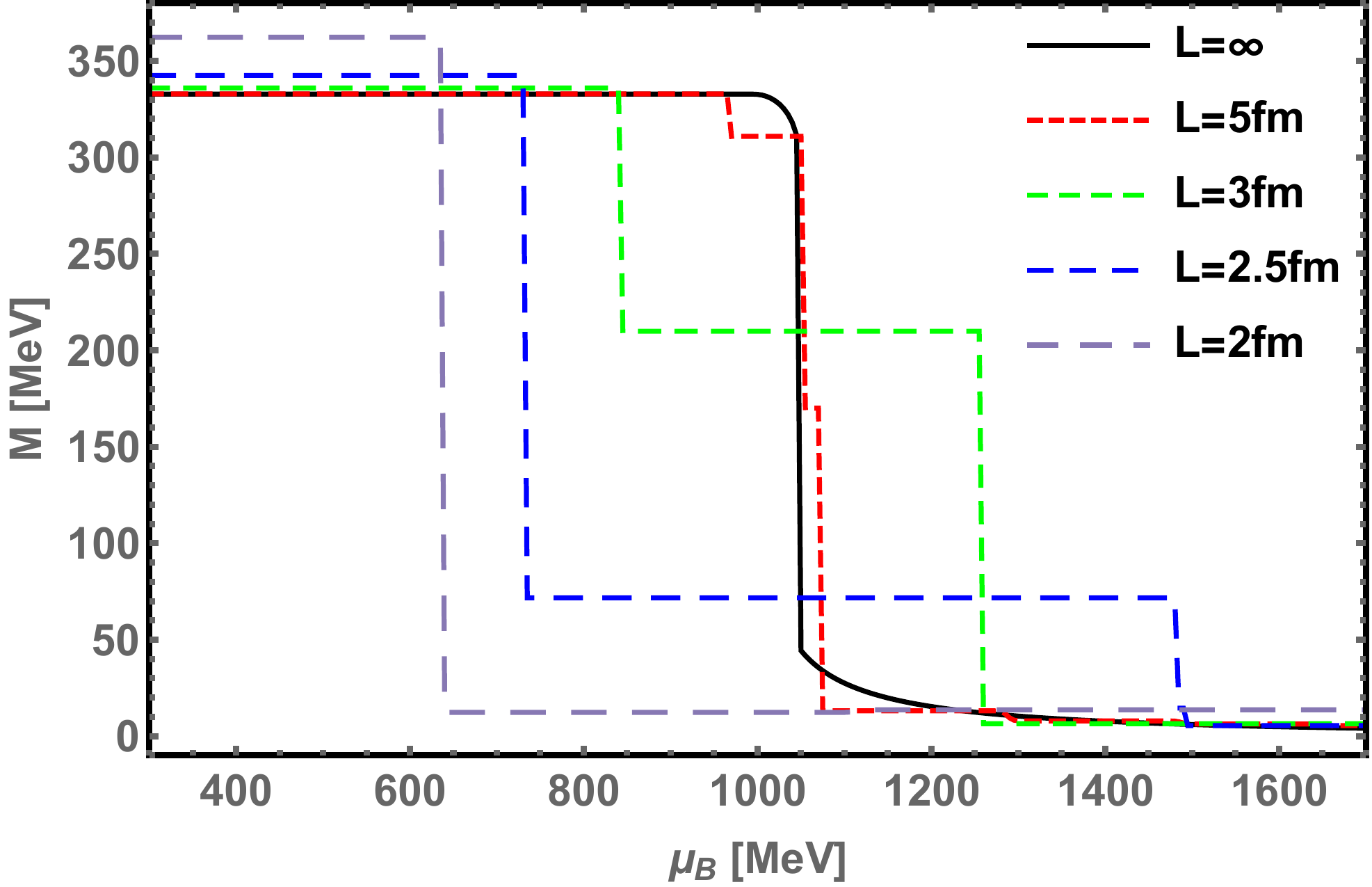}\label{subfig:m-muB-L-PV}\hspace{20pt}
	\caption{Constituent quark mass $M$ as a function of baryon chemical potential for different sizes $L=\infty$ and $L=5,3,2.5,2 fm$ at zero temperature.}
	\label{fig:m-mu-L}
\end{figure}

Correspondingly, the constituent quark mass $M$ as a function of the baryon chemical potential $\mu_B$ at zero temperature $T=0$ is shown in Fig.\ref{fig:m-mu-L}. It is clearly observed that the multi-jump structure of the first-order phase transition, which is called the quantized phase transition,
show up in the size region of $2<L<5 fm$.

\subsection{The splitting of first-order phase transition}

To analyze why the multi-jump structure of the first-order phase transition show up in small system,  in this part we consider the NJL model in hard-cutoff regularization scheme, which is simpler and more transparent than the Pauli-Villars regulization scheme. It is worthy of mentioning that different regularization schemes would not change the qualitative results at finite size. With hard-cutoff, the thermodynamical potential takes the form of :
 \begin{equation}
 \Omega_{\Lambda}=\frac{(M-m_{0})^{2}}{4 G}-\frac{2 N_{c} N_{f}}{V}\sum_{\vec{p}}\left\lbrace E +T\ln(1+e^{-\frac{E+\mu}{T}})+T\ln(1+e^{-\frac{E-\mu}{T}})\right\rbrace ,
 \end{equation}
where the momentum taken into account should be smaller than the cutoff $\Lambda$:
\begin{equation}
\Lambda^{2}>p^{2}=n(\frac{2\pi}{L})^{2},
\label{equ:inequ}
\end{equation}
with $n=\sum_{i=x,y,z}n_{i}^{2}$ and $n_i$ are non-negative integers.

We focus on the chiral phase transition at high baryon chemical potential in the case of zero temperature $T=0$ in this part. We firstly consider the case of small size so that $2\pi/L$ is larger than $\Lambda$, in this case only the zero-momentum mode $n=0$ contributes to the system, and the gap equation Eq.(\ref{gapequ}) is given by:
\begin{equation}
\label{equ:zeromode}
\frac{M-m_{0}}{2G}=\frac{2 N_{c} N_{f}}{V}[1-\theta(\mu-E_{0})],
\end{equation}
where $E_{0}=\sqrt{M^{2}+0(2\pi/L)^{2}}=M$ and $\theta(x)$ is the step function. The solution to this equation is straightforward:
\begin{equation}
M-m_{0}=\left\{
\begin{array}{lr}
	\frac{4GN_c N_f}{V} , \quad\mu<\mu^{c}  & \\
	0\quad\quad ,\quad \mu>\mu^{c} &
\end{array}
\right.
,
\end{equation}
where $\mu^{c}$ is the critical quark chemical potential at zero temperature. Actually, this is a first-order phase transition between the chiral symmetry breaking phase ($\mu <\mu^{c}$) and chiral symmetry restored phase  ($\mu >\mu^{c} $),  just as shown in Fig.\ref{subfig:l2}.


Next we consider a little bit bigger size so that both $n=0$ and $n=1$ can contribute to the system, i.e., both zero-mode and the first-mode are taken into account, and the gap equation becomes:
\begin{equation}
\label{equ:2mode}
\begin{split}
	\frac{M-m_{0}}{2G}=&\frac{2 N_{c} N_{f}}{V}[1-\theta(\mu-E_{0})]+6\frac{2 N_{c} N_{f}}{V}\frac{M}{E_{1}}[1-\theta(\mu-E_{1})],
\end{split}
\end{equation}
where $E_1=\sqrt{M^{2}+(2\pi/L)^{2}}$. The second term in the right-hand side is from the contribution of the first mode and 6 is the degeneracy number of the first mode. The solution to this equation is not as straightforward as Eq.(\ref{equ:zeromode}), however, we can still extract some useful information. Due to the two step functions in the above gap equation Eq.(\ref{equ:2mode}), two phase transitions are expected, and this can be verified from Fig.\ref{subfig:l5} and Fig.\ref{subfig:l3}, and also from the line $L=2.5fm$ and $L=3fm$ in Fig.\ref{fig:m-mu-L}.

In general, more step functions will appear in larger sizes therefore more jumps are expected, however, higher modes don't contribute a lot as long as the size is small. For example, the second term in the right-hand side of Eq.(\ref{equ:2mode}) is small compared to the first term because $M<<E_1$ therefore $M/E_1<<1$ at small size. This also can be verified from Fig.\ref{fig:m-mu-L}: the magnitude of the second jump PT2 at $L=2.5fm$ is smaller than that at $L=3fm$.

\begin{figure}
	\centering
	\includegraphics[width=280pt]{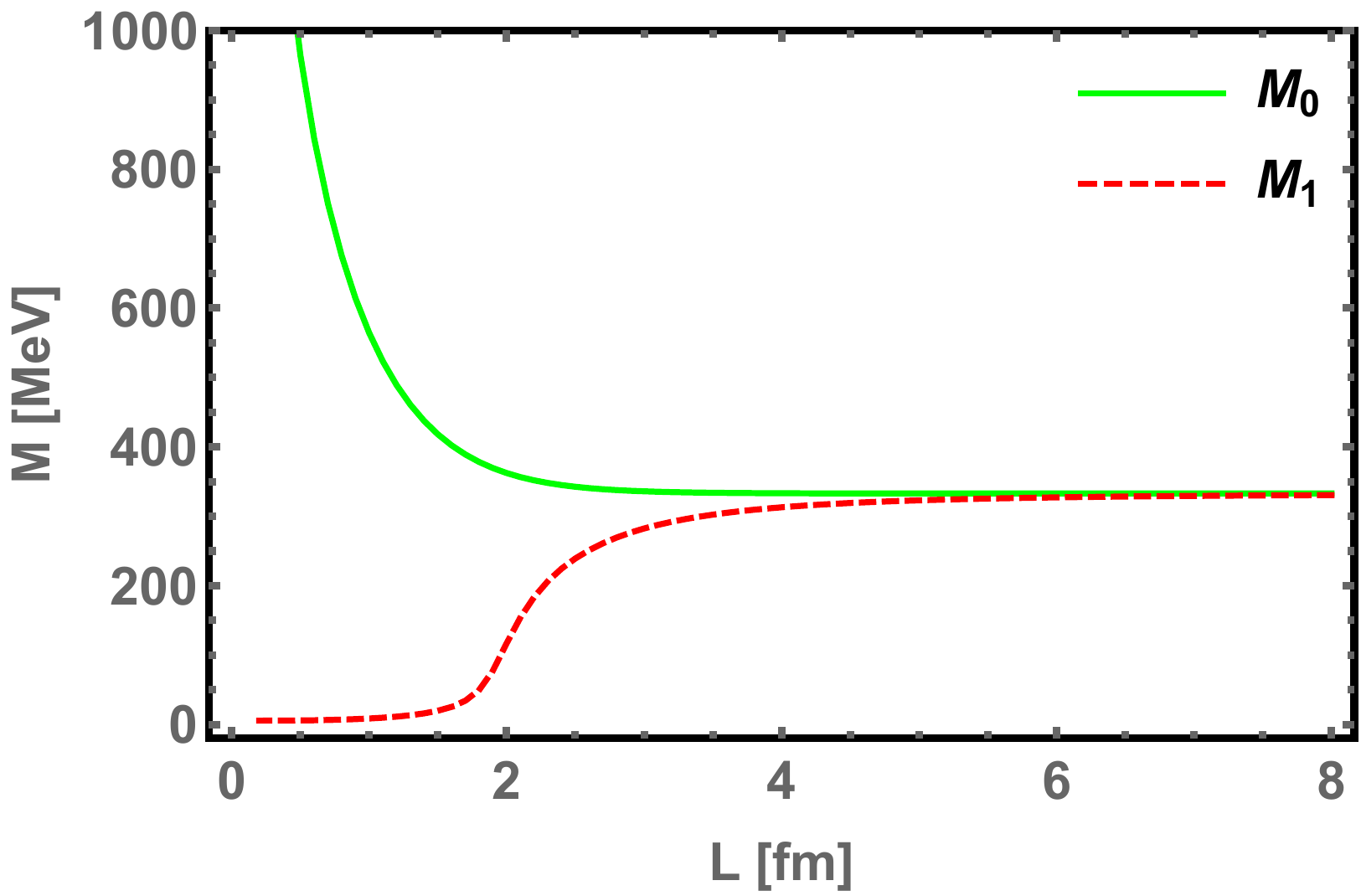}\label{subfig:m0-m1-L-PV}\hspace{20pt}
	\caption{The constituent quark mass $M$ as a function of the size $L$ at zero temperature and zero chemical potential. $M_0$ is obtained by solving gap equation with zero mode while $M_1$ without zero mode.}
	\label{fig:m0-m1-L}
\end{figure}

We show the constituent quark mass $M$ as a function of the size $L$ at zero temperature and zero chemical potential in Fig.\ref{fig:m0-m1-L}. In this plot,  $M_0$ and $M_1$ are obtained by solving the gap equation with and without the zero mode contribution, respectively. At large size when $L>5 fm$, $M_0$ and $M_1$ are almost equivalent, which indicates that the contribution from the zero mode can be ignored at large size. However, when the size decreases, $M_0$ and $M_1$ show completely opposite behaviors: $M_0$ enhances and goes to divergence while $M_1$ decreases and goes to zero at very small size.  From Fig.\ref{fig:m0-m1-L} we can see that the zero-mode contribution becomes dominant at small sizes and there is almost only zero-mode contribution at size $L<2 fm$.

Now we can understand the behavior of quantized first-order phase transition in the size region of $2 fm<L<5 fm$: in general, there are two phase transitions PT1 and PT2, where PT1 is the jump between the chiral symmetry breaking phase with quark mass $M_0$ and the chiral "restoring" phase with quark mass $M_1$,  and PT2 is the jump between the phase with quark mass $M_1$ and the chiral restoration phase with $M=0$, i.e., chiral symmetry totally restored phase. At large size $L>5 fm$, $M_0 =M_1$, which means that the PT1 vanishes, and at small size $M_1 =0$ thus the PT2 vanishes. In conclusion, there is only one jump of the first-order phase transition at both large $L>5 fm$ and small sizes $L<2 fm$, while there appears two jumps for the first-order phase transition in the region of $2fm <L< 5 fm$. This is exactly what we have seen in Fig.\ref{fig:quarkmass-mu-T-L} and Fig.\ref{fig:m-mu-L}.

\subsection{Two sets of critical end point}

\begin{figure}
\centering
\includegraphics[width=280pt]{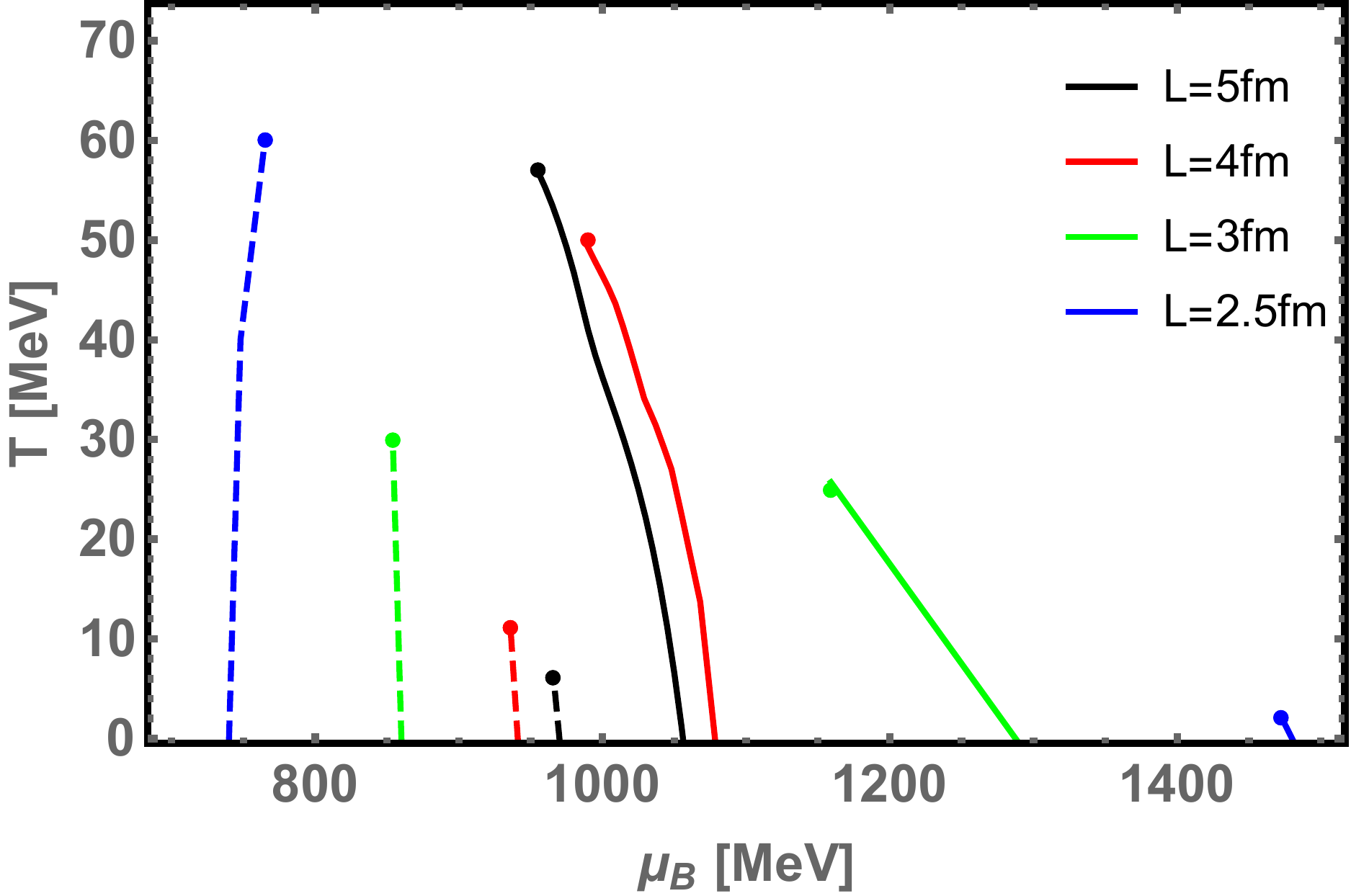}\label{subfig:ceps-PV}\hspace{20pt}
\caption{Two branches of first order phase transitions and two sets of critical end points in the $(T,\mu_B)$ plane for different sizes $L=5,4,3,2.5,2 fm$. The dashed lines and solid lines correspond to PT1 and PT2, respectively.}
\label{fig:ceps}
\end{figure}

Due to the appearance of the two jumps in the first-order phase transition, there are two branches of first-order phase transitions and thus two sets of critical end point showing up in the $(T,\mu_B)$ plane. We show the results in Fig. \ref{fig:ceps} for different sizes $L=5,4,3,2.5,2 fm$, where the solid lines and the CEPs at their ends are corresponding to the PT2 and the dashed lines and the CEPs corresponding to the PT1. We call the CEPs corresponding to PT1 and PT2 as "CEP1"  and "CEP2", respectively. From Fig.\ref{fig:ceps} we can see that CEP1 and CEP2 have opposite behaviors: CEP2 moves to region of higher chemical potential and lower temperature, while CEP1 moves to region of lower chemical potential and higher temperature as size decreases. At last, when the size further decreases, the PT2 and CEP2 disappears and only the PT1 and CEP1 shows up, which are consistent with the behaviors of phase transitions in Fig.\ref{fig:quarkmass-mu-T-L}.

\begin{figure}	
	\centering
	\hspace{-25pt}
	\includegraphics[width=280pt]{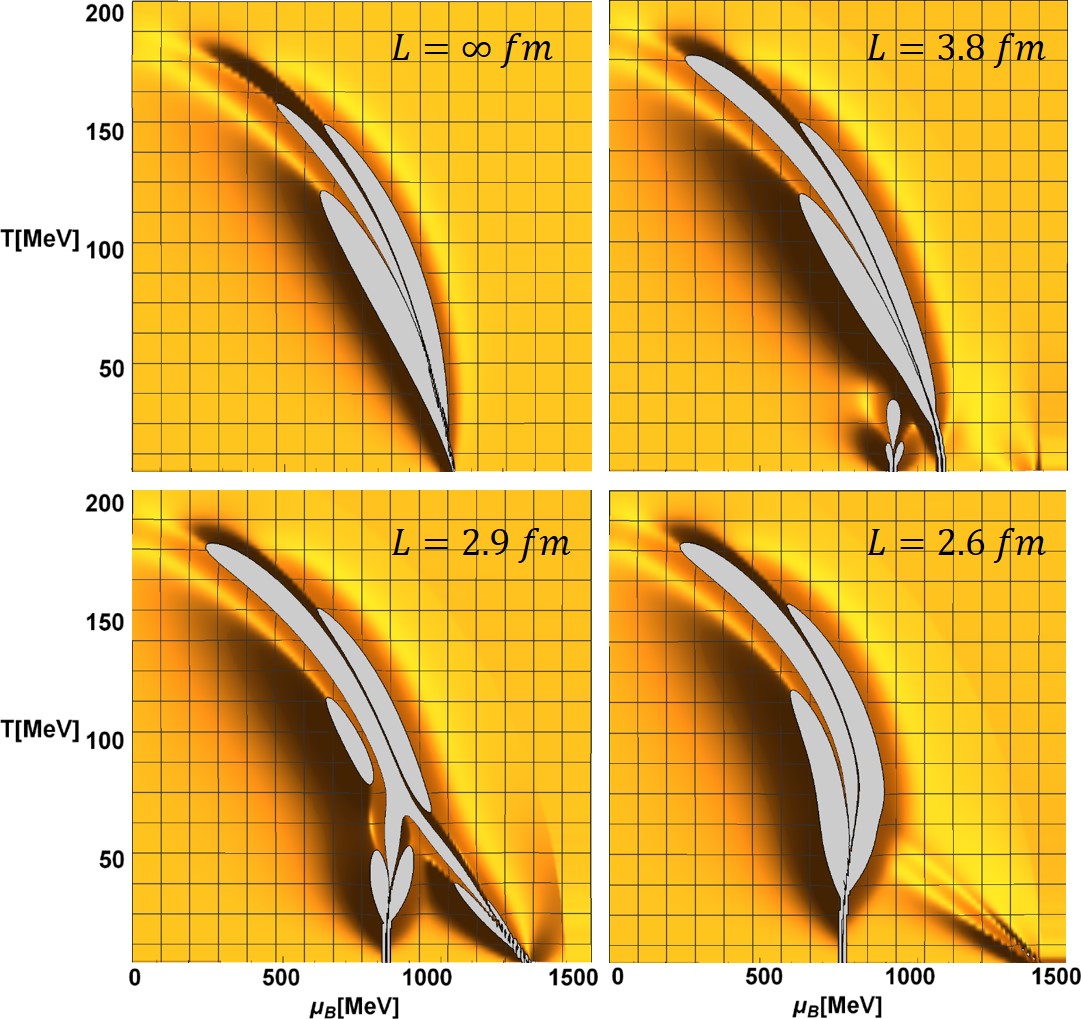}
	\caption{$\kappa \sigma^2$ as functions of the temperature and baryon chemical potential for several different system sizes $L=\infty$ and $L=3.8,2.9,2.6 fm$.}\label{fig:c423d-PV}
\end{figure}

To show more clearly how these two branches of first-order phase transitions evolve with the system size, we show in Fig.\ref{fig:c423d-PV} the 3-dimension (3D) plot of the kurtosis of baryon number fluctuations $\kappa \sigma^{2}$ in the $(T,\mu_B)$ plane. The ratio of the fourth to the second order cumulant of quark number fluctuations is defined as:
\begin{equation}
\kappa \sigma^{2}=\frac{c_{4}}{c_{2}},
\end{equation}
with
\begin{equation}
c_{n}=VT^{3}\frac{\partial^{n}}{\partial(\mu_B/T)^{n}}(\frac{p}{T^{4}}),
\end{equation}
which corresponds to the same ratio for baryon number up to an overall factor $1/9$.  The kurtosis $\kappa \sigma^{2}$ is used as a measurement to locate the CEP in the beam-energy scan at RHIC experiment \cite{Luo:2017faz}. At $L=\infty$, it is clearly seen that there is only one typical first-order phase boundary. When the system size decreases, it is observed that two branches of fist-order phase transition show up in the $(T,\mu_B)$ plane, the branch of PT2 moves to higher chemical potential and lower temperature and eventually disappears, and the other branch PT1 shifts to lower chemical potential and higher temperature region and then becomes dominant.

\begin{figure}	
	\centering
	\hspace{-25pt}
	\includegraphics[width=280pt]{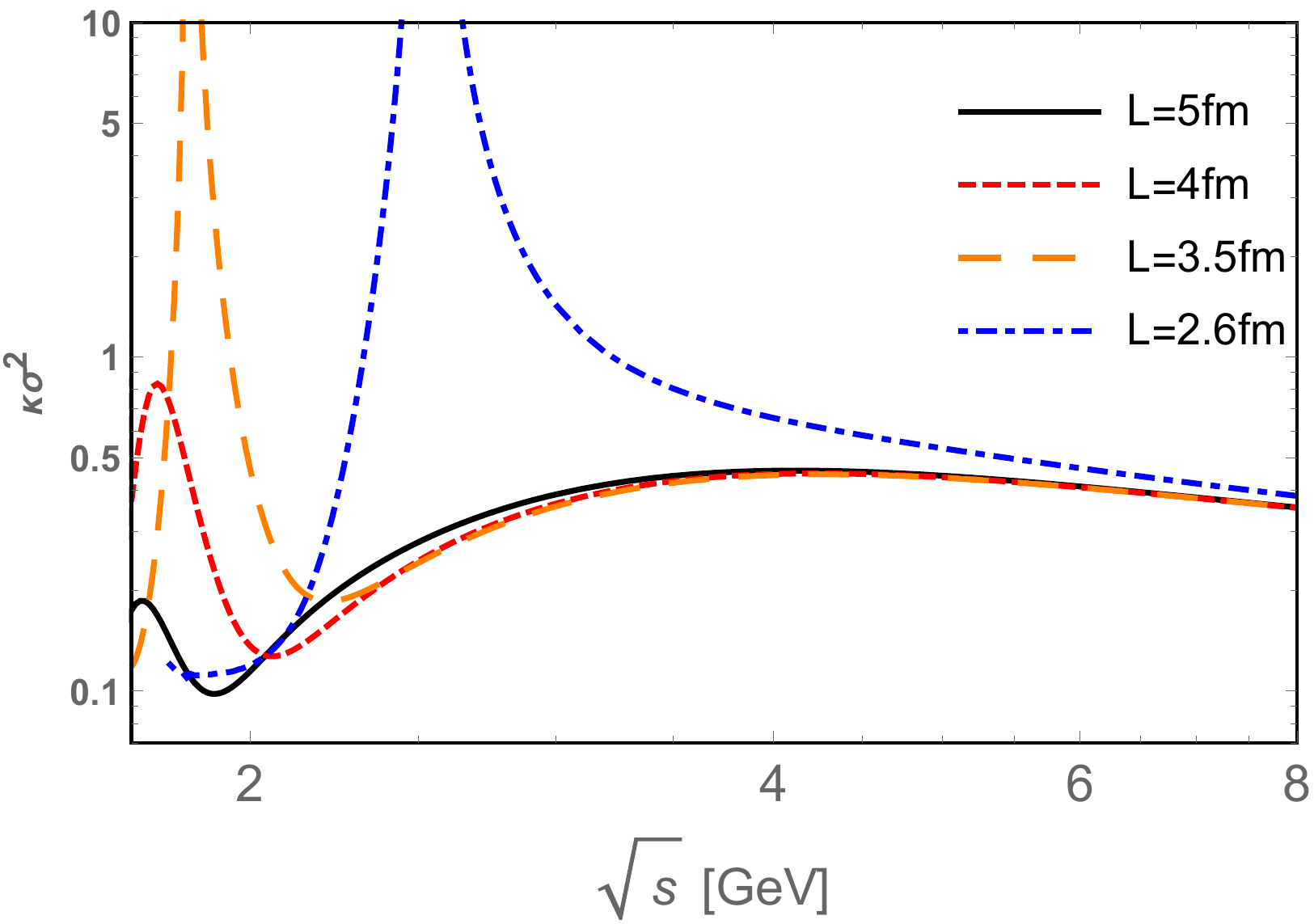}
	\caption{$\kappa \sigma^2$ as a function of the collision energy $\sqrt{s}$ along the experimental freeze-out line for several different system sizes $L=5,4,3.5,2.6 fm$.}\label{fig:kappa-sigma2-ss-L}
\end{figure}

Because the kurtosis $\kappa \sigma^{2}$ is used as a measurement to investigate the existence of the CEP and further to locate the CEP in the beam-energy scan at RHIC experiment, we also show how it would look like if there are two sets of CEPs in the $(T,\mu_B)$ plane. We choose the experimental freeze-out line which is close to the phase boundary, and show the kurtosis $\kappa \sigma^{2}$ along the freeze-out line as a function of the collision energy $\sqrt{s}$ in Fig. \ref{fig:kappa-sigma2-ss-L} for different size, where the relation between the collision energy $\sqrt{s}$ and the baryon chemical potential$\mu_{B}$ is\cite{Luo:2017faz}:
\begin{equation}
\mu_B(\sqrt{s})=\frac{1.477}{1+0.343\sqrt{s}}.
\label{equ:mu-ss}
\end{equation}
From Fig.\ref{fig:kappa-sigma2-ss-L}, we can see that there will be double-peak structure of the $\kappa \sigma^{2}$  showing up along the freeze-out line, which is fitted to experimental data \cite{Luo:2017faz}:
\begin{equation}
T(\mu_B)=0.158-0.14\mu_B^{2}-0.04\mu_B^{4}.
\label{equ:fo}
\end{equation}
Be aware of that the unit of $T$ and $\mu_B$ in Eq.(\ref{equ:mu-ss}) and Eq.(\ref{equ:fo}) is GeV.
\section{Summary}

The finite size effect on hadron physics and quark matter of QCD has attracted much interest for more than three decades, however, there still exists the ambiguity of applying the boundary conditions for quarks, i.e. whether to apply the periodic or the anti-periodic spatial boundary condition. In this talk, we consider the NJL model in a finite volume. To take into account the finite size effect, we replace the momentum integral by momentum summations.

 By comparing the thermodynamical potential of quark matter at finite size, it is found that the ground state of small system of quark matter favors the periodic spatial boundary condition for quarks. In the stable small system with periodic boundary condition, it is observed that the chiral symmetry breaking enhances in the vacuum, which is called the catalysis of chiral symmetry breaking, and the pions excited from the droplet vacuum keep as pseudo Nambu-Goldstone bosons and pion mass keeps as a constant in finite size system. The phenomena of the catalysis of chiral symmetry breaking as well as pseudo Nambu-Goldstone pions in small system are similar to those in quark matter under strong magnetic fields. The similarity between these two systems is understandable if we remember that the magnetic length $l$ for particle carrying charge $q$ is proportional to the inverse of the square root of magnetic field, i.e. $ l\sim \frac{1}{\sqrt{|q| B}}$. In some sense, we can imagine the system of charged particles under strong magnetic field as putting these particles in an elongated cylinder with small radius $l$.

 The only difference between the periodic and anti-periodic boundary conditions is that the momentum summation starts from the exact zero-momentum in periodic boundary condition thus the zero-momentum mode is taken into account in the periodic boundary condition. From the constituent quark mass at finite size, we can see that the contribution from the zero-momentum mode becomes dominant in small size system.

More interestingly, the zero-momentum mode brings significant change on the chiral phase transition in small system of cold dense quark matter. It is found that in some region of size $2fm <L<5 fm$, the first-order chiral phase transition becomes quantized, and there are two branches of first-order phase transitions and thus two sets of critical end point showing up in the $(T,\mu)$ plane. It is worthy of mentioning that this is the first time to observe the quantized first-order phase transition in literatures, which is a brand new phenomena. The quantized first-order phase transition is induced by the quantized momentum summation and in the size region when the zero-momentum mode contribution becomes dominant. Similar phenomena is also observed in quark matter under strong magnetic field \cite{Xukun-long}, and it can be expected that such phenomena can show up in some small systems in condensed matter, and in mini black holes.

At last, it is worthy of mentioning that in this work, we didn't consider the gluon dynamics in finite system, which may affect the magnitude of interaction in the scalar channel between quarks thus affect the properties of quark matter in finite system. We leave this for future studies.



\begin{acknowledgments}
We thank valuable discussions with T.Hatsuda and R.Pisarski. This work is supported in part by the NSFC under Grant Nos. 11725523, 11735007, 11261130311 (CRC 110 by DFG and NSFC), Chinese Academy of Sciences under Grant No. XDPB09, and the start-up funding from University of Chinese Academy of Sciences(UCAS).
\end{acknowledgments}

\end{document}